\documentclass[3p,times,procedia]{elsarticle}

\usepackage{ecrc}

\volume{00}

\firstpage{1}

\journalname{Physics Procedia}

\runauth{}

\jid{phpro}

\jnltitlelogo{Physics Procedia}

\CopyrightLine{2011}{Published by Elsevier Ltd.}

\usepackage{amssymb}

\usepackage[figuresright]{rotating}

\begin{document}

\begin{frontmatter}



\dochead{}
%% e.g. \dochead{17th International Conference on Dynamical Processes in Excited States of Solids}

\title{Detector Systems at CLIC}


\author{Frank Simon}

\address{Max-Planck-Institut f\"ur Physik, Munich, Germany\\
Excellence Cluster `Universe', Garching, Germany}

\begin{abstract}
The Compact Linear Collider CLIC is designed to deliver $e^+e^-$ collisions at a center of mass energy of up to 3 TeV. The detector systems at this collider have to provide highly efficient tracking and excellent jet energy resolution and hermeticity for multi-TeV final states with multiple jets and leptons. In addition, the detector systems have to be capable of distinguishing physics events from large beam-induced background at a crossing frequency of 2 GHz. Like for the detector concepts at the ILC, CLIC detectors are based on event reconstruction using particle flow algorithms. The two detector concepts for the ILC, ILD and SID, were adapted for CLIC using calorimeters with dense absorbers limiting leakage through increased compactness, as well as modified forward and vertex detector geometries and precise time stamping to cope with increased background levels. The overall detector concepts for CLIC are presented, with particular emphasis on the main detector and engineering challenges, such as: the ultra-thin vertex detector with high resolution and fast time-stamping, hadronic calorimetry using tungsten absorbers, and event reconstruction techniques related to particle flow algorithms and beam background suppression. 
\end{abstract}

%\begin{keyword}



%\end{keyword}

\end{frontmatter}

% \linenumbers

\section{Introduction - Experimental Conditions at CLIC}
\label{}

The Compact Linear Collider CLIC \cite{Assmann:2000hg} is a high-energy $e^+e^-$ collider project under development, designed for the exploration and understanding of physics beyond the Standard Model in the TeV energy range. The accelerator concept is a  based on a two-beam acceleration scheme using normal-conducting, high-frequency cavities, capable of reaching a center of mass energy of 3 TeV. CLIC will operate with bunch trains with a length of 156 ns, inter-bunch spacing of 0.5 ns and a train repetition rate of 50 Hz. The luminosity of  $5.9\, \times\, 10^{34}$\, cm$^{-2}$s$^{-1}$ requires very strong focusing of the beams at the collision point, leading, together with the high beam energy, to massive beamstrahlung effects, with 28\% of the energy per bunch lost through radiation effects at 3 TeV. This leads to a collision energy spectrum with 35\% of the total luminosity in the top 1\% of the energy, with a long tail to smaller energies, and results in the abundant creation of coherent and incoherent $e^+e^-$ pairs \cite{Chen:1992ax}. The high luminosity  and high energy also results in copious production of hadronic mini-jets from two photon processes \cite{Drees:1991zka, Chen:561345}.  

\begin{figure}[hbt]
\centering
 \includegraphics[width=0.49\linewidth]{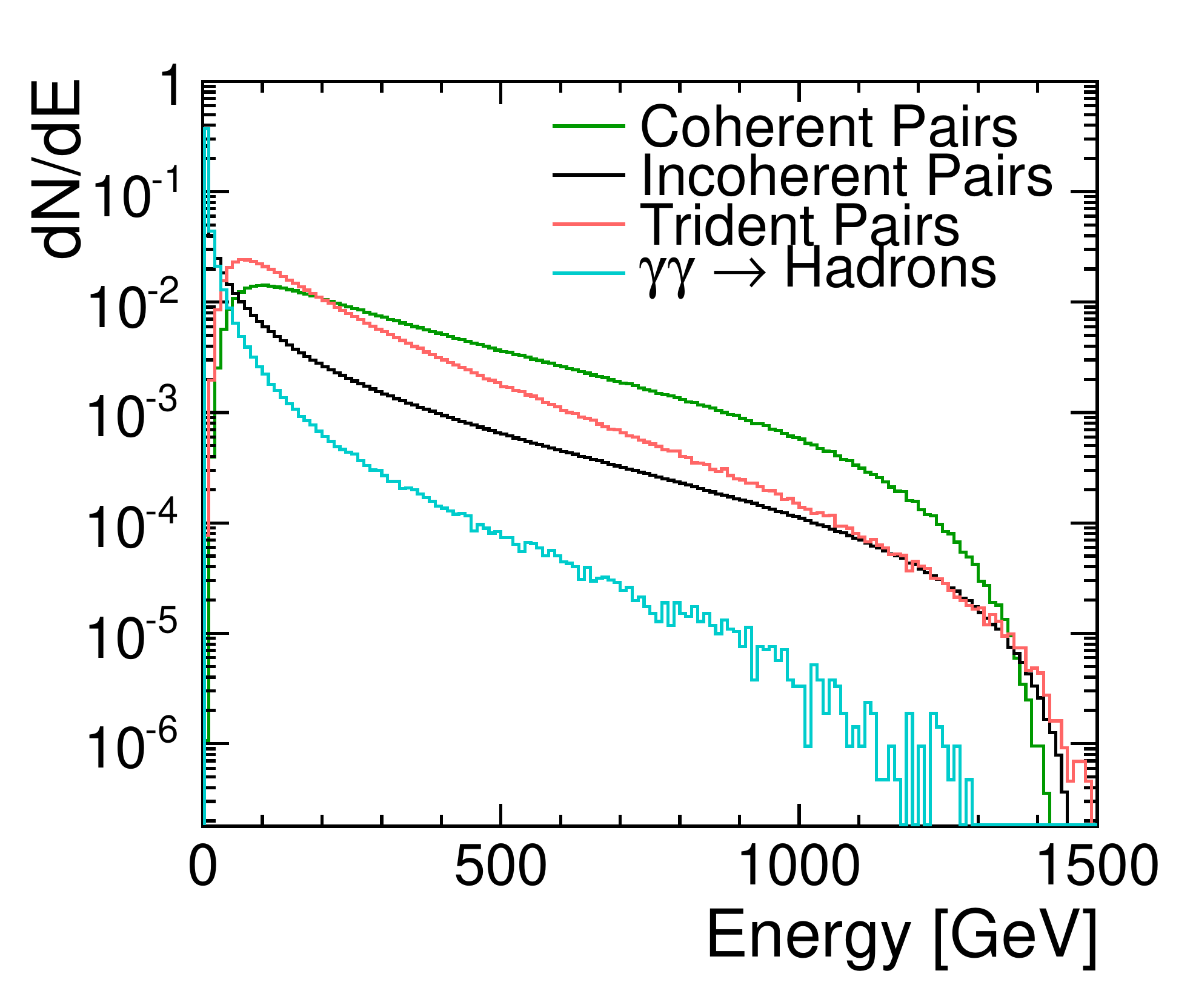} 
 \includegraphics[width=0.49\linewidth]{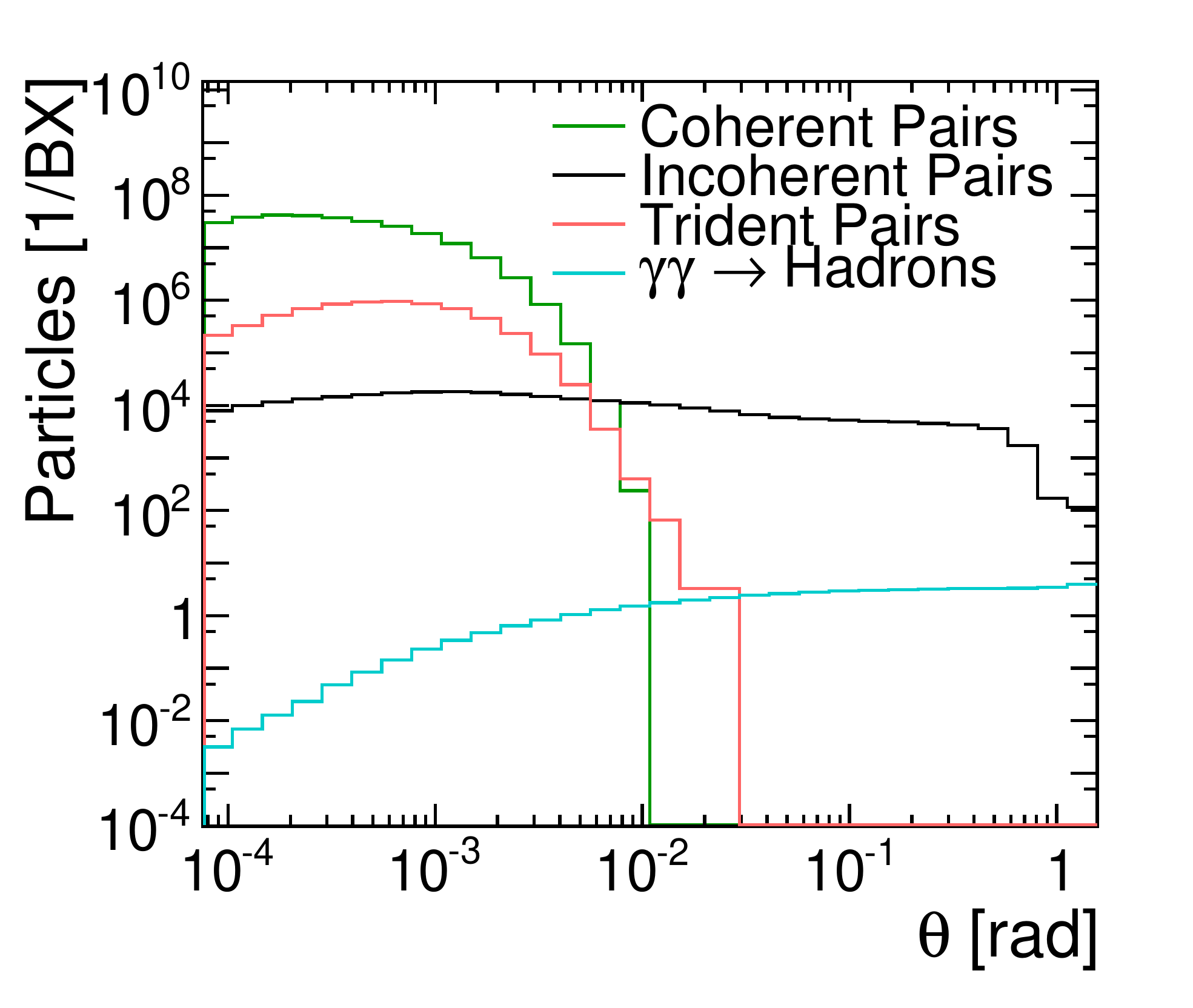}
 \caption{The distributions of the beam related backgrounds: (left) Fraction of
   energies for the particles of each background source. (right)
   Angular distribution of the produced background particles. Both plots are for
   CLIC at 3 TeV.\label{fig:Backgrounds}}
\end{figure}

This environment places strict constraints on the detector design, on the capabilities of individual subsystems, and on the event reconstruction and data analysis. For example, due to the high bunch-crossing frequency of 2 GHz, the  background leads to pileup in the detector, in particular in the case of $\gamma\gamma \to  {\rm hadrons}$ processes, requiring excellent time stamping in the detector systems.  The coherent pairs, with a total of  $2.1 \,\times\, 10^8$ TeV per bunch crossing for 3 TeV collisions, are emitted at small opening angles of a few milliradian, requiring a beam crossing angle of 20 mrad. The incoherent pairs, which have significantly larger angles, put constraints on the dimension of the beam pipe and on the position of the inner tracking elements. Figure \ref{fig:Backgrounds} shows the energy distribution and the angular distribution of background particles from different processes. Only incoherent pairs and hadronic backgrounds reach significantly into the detector acceptance.

\section{CLIC Detector Concepts}

The physics goals of CLIC require detector systems with excellent track and jet reconstruction, very low mass trackers, highly efficient flavor tagging and particle identification and a hermetic coverage. These requirements are essentially identical to those for detectors at the International Linear Collider ILC \cite{Brau:2007zza}, albeit at a lower energy. For the ILC, two general-purpose detector concepts, ILD \cite{ildloi:2009} and SiD \cite{Aihara:2009ad}, have been developed over the last decade and have been thoroughly evaluated \cite{IDAG2009}. These concepts are used as starting points for CLIC detectors, with modifications motivated by the more challenging experimental conditions at CLIC and by the higher collision energy, and are referred to as CLIC\_ILD and CLIC\_SiD.

\begin{figure}[!htb] 
	\centering

\includegraphics[scale=0.72,clip,trim=0 16.5 0 20]{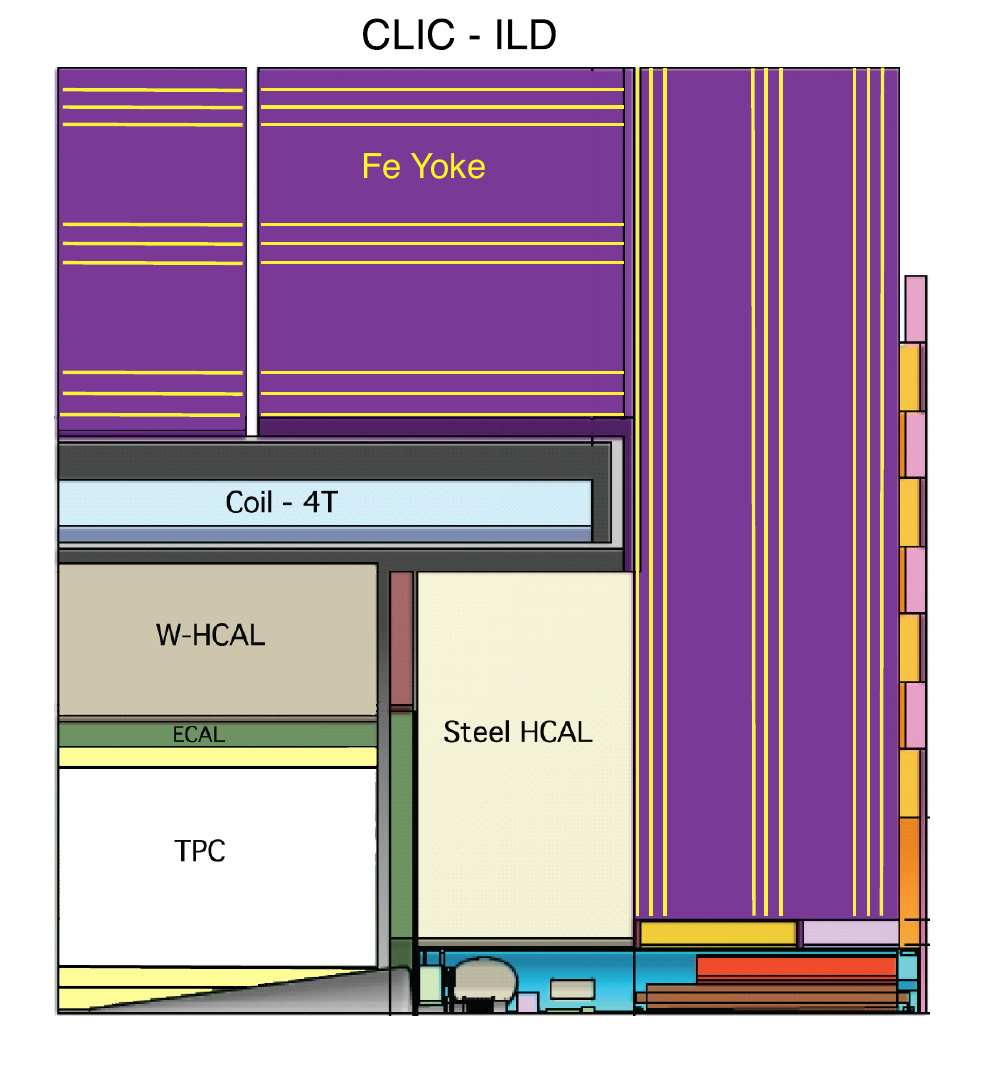}
\hfill
\includegraphics[scale=0.46,clip,trim=0 0 0 20]{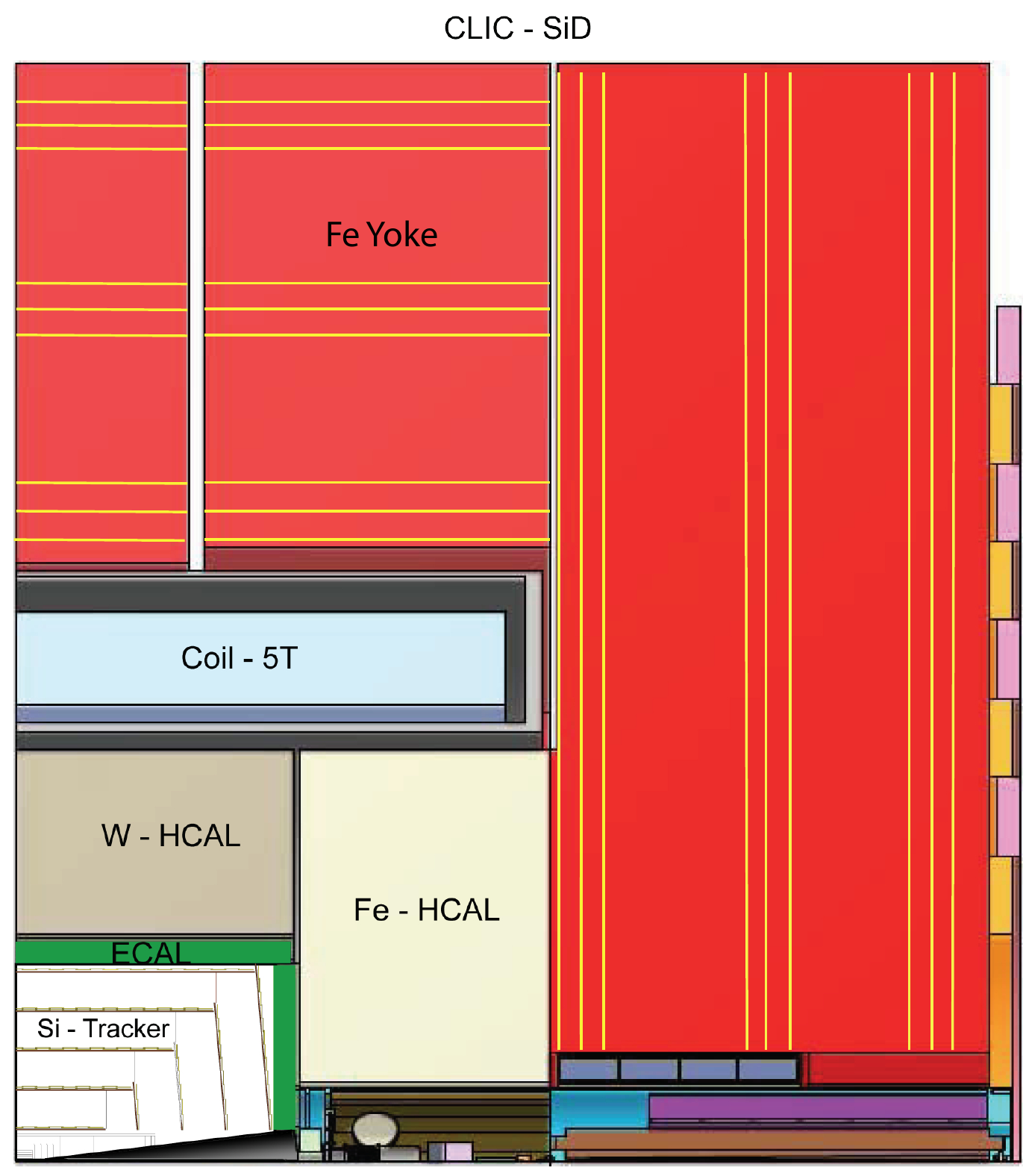}

	\caption{Longitudinal cross section of top quadrant of CLIC\_ILD and CLIC\_SiD, to scale. }
	\label{fig:DetectorView}
\end{figure}

Both of these concepts have a similar general design, with a large solenoid which contains a tracking system and electromagnetic and hadronic calorimetry. They are optimized for particle flow event reconstruction \cite{pfaMorgunov, Brient:2002gh, thomson:pandora}, with efficient precision tracking and highly granular calorimetry. However, they differ in the detailed approach. CLIC\_ILD aims for an optimized jet reconstruction with calorimetry at a relatively large radius and consequently a smaller magnetic field of 4 T, and a large TPC as main tracking detector. CLIC\_SiD is designed as a compact, cost-optimized detector with a 5 T solenoid and high precision all-silicon tracking. As for the ILC, CLIC is foreseeing a push-pull mechanism to alternately operate both detectors at a single interaction region.

Figure \ref{fig:DetectorView} shows the overall layout of the two detector concepts. Due to the higher magnetic field and consequently a thicker return yoke of CLIC\_SiD, both detectors have the same overall radius of 7 m. To facilitate push-pull operations, both concepts also have the same length of 12.8 m. The reduction of stray fields from the solenoids requires additional compensation coils on the outside of the detector endcaps in the case of CLIC\_ILD.

There are two main challenges at CLIC that require substantial changes of the detector concepts compared to those of the ILC. These are the higher collision energy, and the high rate of background from $\gamma\gamma \, \to \, {\rm hadrons}$ and incoherent $e^+e^-$ pairs.  While the high energy affects the design of the calorimeter system, the high background rate requires time stamping on the nanosecond level in some and on the 10 ns level in most detector systems, and an increase of the inner radius of the vertex detector. In the following, the main detector subsystems are briefly described. 

\subsection{Vertex Detector}

\begin{figure}[!ht] 
	\centering
\includegraphics[width=0.99\textwidth]{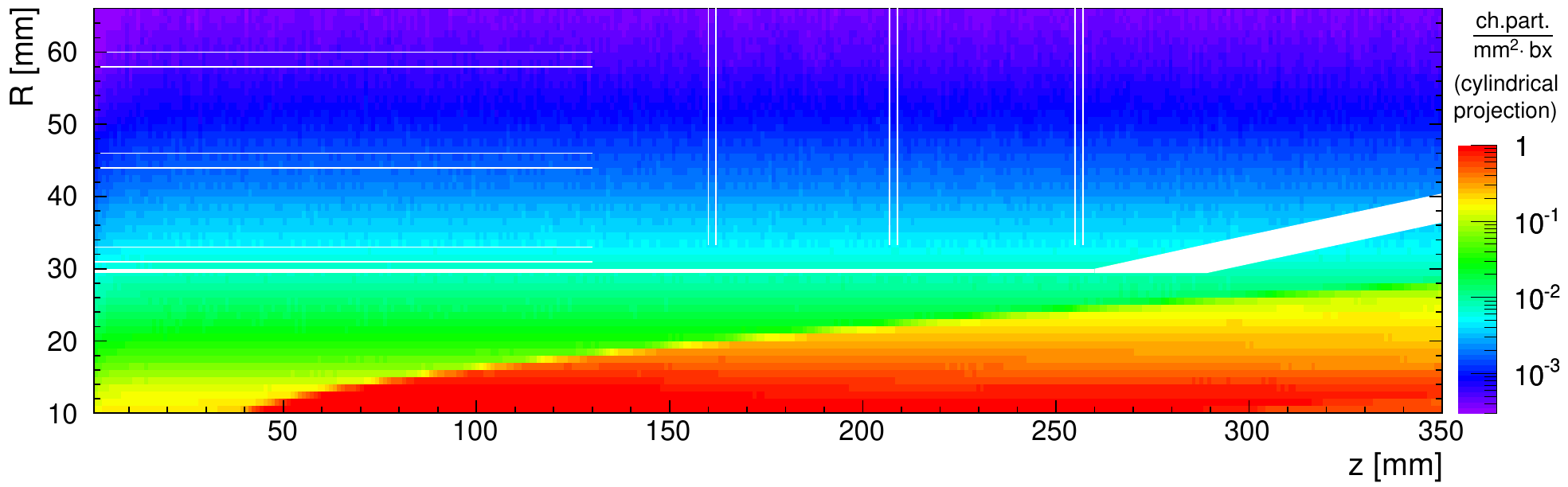}
\caption{Density of direct hits from incoherent pairs in the vertex region of CLIC\_ILD. The beam pipe as well as the positions of the vertex detector layers are indicated by white lines}.
\label{fig:VertexBackground}
\end{figure}

The vertex detector, a silicon pixel tracker, is located as close as possible to the interaction point to provide optimal secondary vertex reconstruction and accurate flavor tagging. The position of the innermost tracking layer, and the minimum radius of the beam pipe in the interaction region, is defined by the density of incoherent $e^+e^-$ pairs produced from beamstrahlung. Since the magnetic field is higher in CLIC\_SiD than in  CLIC\_ILD, the beam pipe radius is  24.5 mm in the former and 29.4 mm in the latter concept. Figure \ref{fig:VertexBackground} shows the density of direct hits from incoherent pairs in the vertex region of CLIC\_ILD. The innermost layer of the vertex detector is located such that the hit occupancy is less than $\sim10^{-2}$ hits/mm$^2$/ns.

The detailed layouts of the vertex tracker differs between CLIC\_ILD and CLIC\_SiD. While the former uses a barrel detector with 3 double layers and three double forward disks, the latter uses 5 single layers in the barrel and a total of 7 forward disks. While the double layers allow a construction with a lower material budget and thus reduced multiple scattering, the 5 single layers add information for the track pattern recognition, which is important in the CLIC\_SiD case due to the smaller number of hits in the all-silicon tracking system. 

To satisfy the requirements for heavy-flavor and tau identification, a transverse impact parameter resolution of $(\sigma(d_0))^2=a^2+b^2\cdot\mathrm{GeV}^2/(p^2 \mathrm{sin}^3\theta)$, with $a  \approx 5\  {\rm \mu m}$ and $b  \approx 15\  {\rm \mu m}$ is needed. Simulations have shown that this can be achieved with single-point resolutions of approximately 3 $\mu$m and a material budget of less than 0.2\% $X_0$ for the beam pipe and each of the tracking layers, respectively. The technology for the vertex detector thus has to make use of thin silicon with pixel sizes in the 20 $\mu$m range and analog readout. In addition, a time stamping on the 10 ns level is desirable to limit the impact of the high pair background on the tracking. Several technological options are being considered, such as thinned hybrid pixels with 3D interconnects, active pixel sensors, or fully 3D integrated solutions. To keep the power consumption and thus the cooling needs to a minimum, power-pulsing at the bunch train frequency of 50 Hz will be essential. 

\subsection{Main Tracker}

The main trackers of CLIC\_ILD and CLIC\_SiD follow very different approaches. In both cases, the designs of the ILC detector concepts have been adopted. 

For CLIC\_ILD, the main tracking is provided by a large TPC with a radius of 1.8 m and a length of \mbox{4.5 m} with micro-pattern gas detector readout. It is complemented by silicon tracking layers outside of the TPC, which are necessary to achieve excellent momentum resolution, and to provide the time stamping capability necessary to perform reliable tracking in the environment of high backgrounds. 

CLIC\_SiD uses a five layer silicon strip tracker with a barrel and an endcap section, with a total radius of 1.2 m and a length of 3.3 m. The tracking layers in the barrel are constructed from 10 $\times$ 10 cm$^2$ silicon modules with 300 $\mu$m sensitive thickness and 25 $\mu$m wide strips, with every second strip read out, resulting in a readout pitch of 50 $\mu$m. In the endcaps, trapezoidal silicon modules with the same readout pitch and similar overall size, mounted as two stereo layers per tracking layer, will be used. 

Detailed simulation studies have shown that both tracking systems achieve the target resolution of \mbox{$2\cdot10^{-5}$ GeV$^{-1}$} for single high-energy muons, with CLIC\_SiD exceeding this goal for particles in the central region.

\subsection{Calorimetry}

Since both CLIC detector concepts are based on particle flow event reconstruction, the calorimeters are of particular importance. 

In the electromagnetic calorimeters, the separation of close-by electromagnetic showers is crucial, requiring dense absorbers and a lateral and longitudinal segmentation below the Moli\`ere radius $R_M$ and below one radiation length $X_0$, respectively. The ECAL designs for the CLIC detectors are taken over from the ILC detectors without modifications. Both concepts use silicon-tungsten electromagnetic calorimeters as baseline technology, with active elements with a pad size of approximately 25 mm$^2$ in the case of CLIC\_ILD and of 13 mm$^2$ in the case of CLIC\_SiD. Both concepts use 30 layers, with the first 20 layers with a twice finer sampling, and a total thickness of 23 $X_0$ and 26 $X_0$ for CLIC\_ILD and CLIC\_SiD, respectively. In addition to the silicon-tungsten option, the use of small scintillator strips with SiPM readout as active medium is being considered, as well as mixed designs using alternating layers of silicon and scintillator. 

The high energy at CLIC, with jets up to the TeV region, places particular requirements on the depth of the hadronic calorimeter systems. At the same time, the overall thickness of the calorimeter is limited by the available free bore of the solenoid, which can not be substantially increased compared to the ILC detectors due to technological limitations and cost. It is thus mandatory to build the calorimeters as compact as possible, while providing a large overall interaction length. This led to the choice of tungsten as absorber material for the barrel hadronic calorimeter.

\begin{figure}[!ht] 
	\centering
\includegraphics[width=0.49\textwidth]{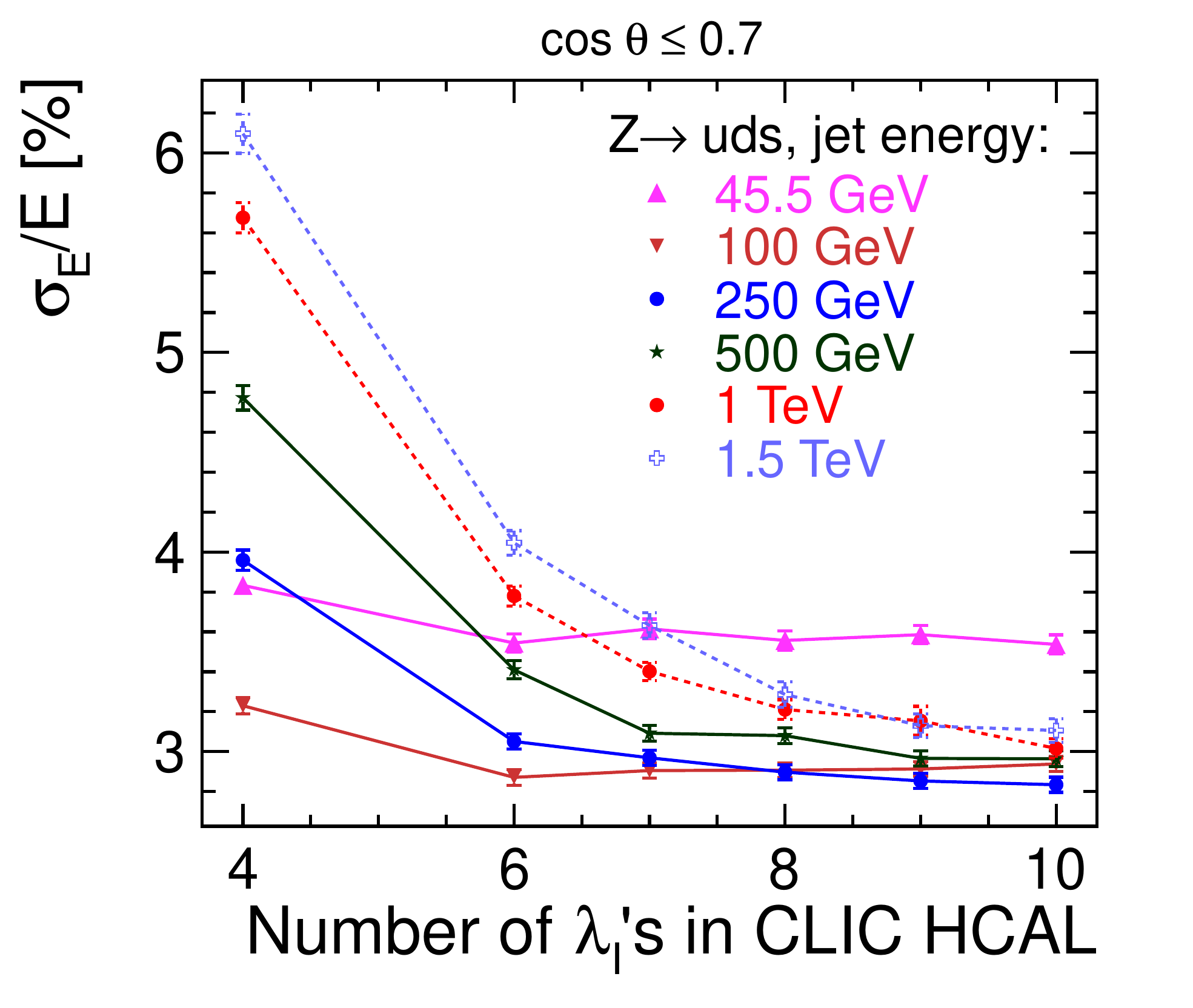}
\includegraphics[width=0.49\textwidth]{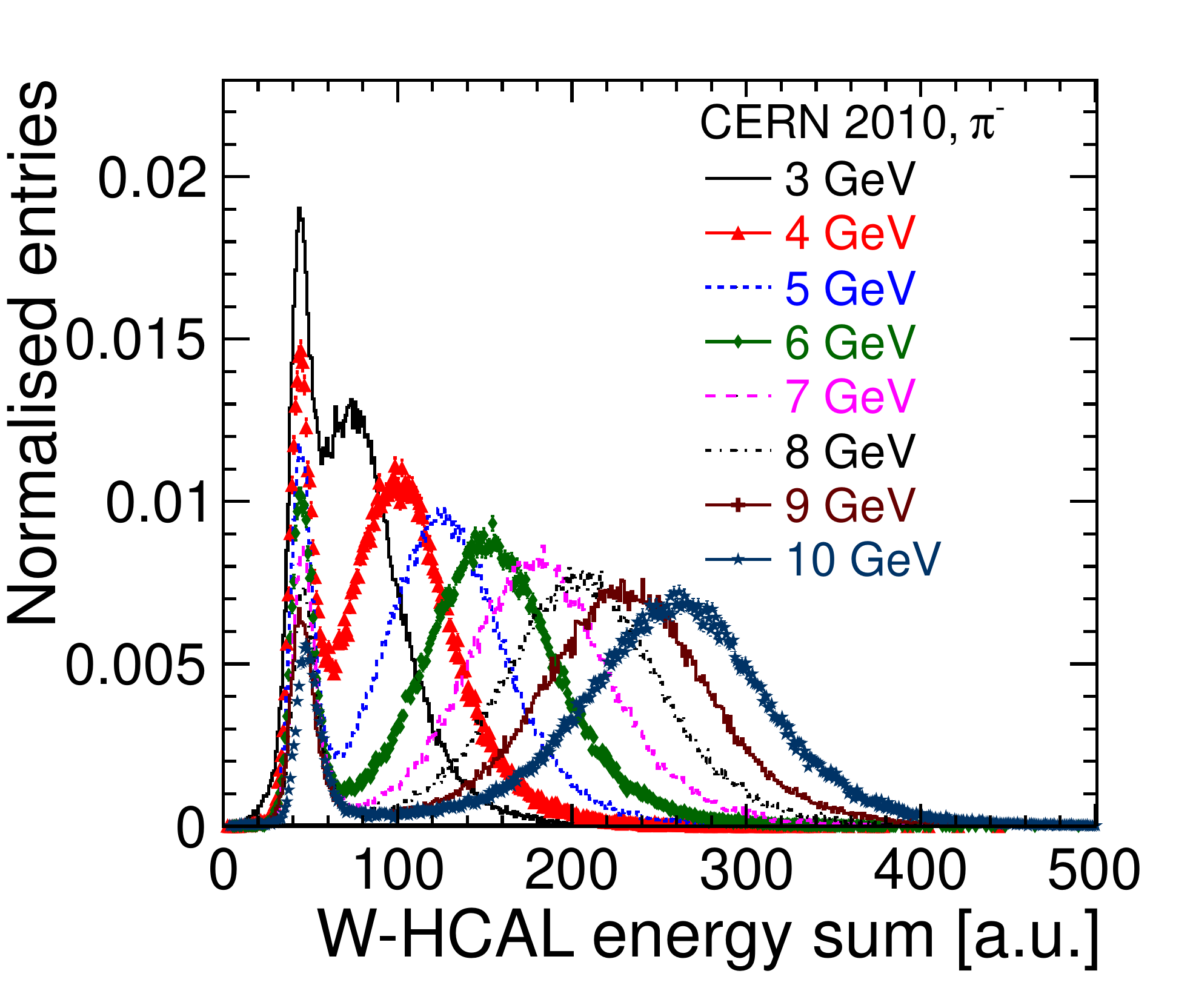}
\caption{Left: Simulated jet energy resolution for various jet energies using particle flow with a tungsten HCAL in the CLIC\_ILD concept as a function of the calorimeter depth.Right: Response of the tungsten-scintillator calorimeter prototype to low-energy pions. The low-energy peak, present in all distributions, is due to the muon contamination of the beam.}.
\label{fig:CalorimeterDepth}
\end{figure}

Figure \ref{fig:CalorimeterDepth} (left) shows the jet energy resolution for various energies using particle flow event reconstruction in the CLIC\_ILD concept as a function of the depth of a tungsten HCAL with scintillator readout. In front of the HCAL, there is a 1 $\lambda_I$ deep silicon-tungsten ECAL. This study suggests the choice of a 7.5 $\lambda_I$ deep hadronic calorimeter, since the resolution quickly degrades for thinner systems, while only moderate improvement can be achieved with additional depth. For the endcap calorimeters, the space constraints are significantly relaxed, allowing the use of steel as absorber medium, which is considerably less expensive and provides a fast shower development, which is important in this region of higher backgrounds. For the current simulation studies in the framework of the CLIC CDR, active elements with small plastic scintillator tiles with embedded SiPMs are used. For the HCAL, several alternative readout technologies based on gaseous detectors are also studied within the framework of the CALICE collaboration, such as RPCs with digital or semi-digital readout as well as Micromegas and GEM detectors.  

%\begin{figure}[!ht] 
%	\centering
%\includegraphics[width=0.5\textwidth]{figures/WHCALEnergy.pdf}
%\caption{Response of the tungsten-scintillator calorimeter prototype to low-energy pions. For the low-energy peak, present in all distributions, is due to the muon contamination of the beam.}.
%\label{fig:WHCALEnergy}
%\end{figure}

To study the performance of a hadron calorimeter with tungsten absorbers, a 38 layer prototype using the scintillator layers of the CALICE analog hadron calorimeter \cite{Adloff:2010hb} has been constructed and tested extensively in particle beams over an energy range from 1 GeV to 300 GeV. The data analysis is still in an early phase and indicates a good performance of the detector for hadron beams. Figure \ref{fig:CalorimeterDepth} (right) shows the response of the prototype calorimeter to low energy pions. 

The calorimeters play a crucial role in the rejection of pile-up from $\gamma\gamma \to  {\rm hadrons}$ background. Nanosecond time resolution on the cluster level is expected in both the ECAL and the HCAL. In addition to the time stamping itself, the required integration time also plays an important role. Here, significant differences between a steel and tungsten HCAL are expected, due to the increased importance of slow shower components such as nuclear fragments and neutron-induced processes in heavier absorbers. First studies of the time structure of hadronic showers in a scintillator tungsten HCAL have been performed in the calorimeter testbeam, demonstrating the importance of correct neutron treatment in the shower simulation \cite{SimonT3B}. 

\subsection{Magnet System and Instrumented Return Yoke}

The magnet system for the CLIC detectors consists of the main solenoid and two forward anti-solenoids around the final focusing elements. In the case of CLIC\_ILD, additional compensation coils are located on the endcaps to satisfy the stray field requirements despite a thinner yoke, driven by the goal to achieve an equal length for both detector concepts to simplify ``push-pull'' operations . The central solenoids for both concepts have similar parameters, with CLIC\_SiD being the more challenging one due to the higher field of 5 T. The design is based on the experience from the construction of the LHC detector magnets, in particular the CMS solenoid \cite{CMS:2008zzk} and the ATLAS central solenoid \cite{Yamamoto:2008zze}, going beyond some of the parameters, in particular with an energy to mass ratio of 15 kJ/kg. The coil consists of five layers, with a conductor cross section of 97.4 mm $\times$ 15.6 mm. The core of the conductor is a 40-strand NbTi/Cu Rutherford cable with mechanical reinforcement. The latter can either be based on existing materials or on an even stronger new material to be developed.

The flux return for the solenoid, constructed from steel, is instrumented with muon detectors. Due to the large amount of material in front of the return yoke, these detectors can not contribute to the momentum resolution of muons, but they are crucial for the identification of muons. In addition, the first layers serve as a tail catcher for the calorimeter system to improve the energy resolution for late-starting showers that leak beyond the solenoid.  In the endcap regions, the muon system also plays an important role in the identification of beam-halo muons. Together, these different applications for the muon detectors lead to requirements on the readout granularity and on the timing, which are similar to those for the hadronic calorimetry with cell sizes of $30\times 30$ mm$^2$ and a time resolution of 1 ns. 

As shown in Figure \ref{fig:DetectorView}, the muon detectors are arranged in three groups of three layers each. This geometrical arrangement is beneficial for the mechanical stability of the return yoke, while studies have shown that it incurs no performance penalties compared to a uniform spacing of the detector layers throughout the yoke volume. In the barrel region, where the issue of hadronic shower leakage is most critical, the first detector layer is located in between the cryostat of the solenoid and the return yoke, while in the endcap, all layers are embedded in the yoke material. Due to the higher occupancy in the endcap regions from beam halo muons, a pad readout is mandatory there, while in the barrel region also crossed strips are possible. The readout technologies being considered are plastic scintillators with SiPM readout and RPCs.

\subsection{Interaction Region and Detector Integration}

\begin{figure}[htb] 
	\centering
\includegraphics[width=0.9\textwidth]{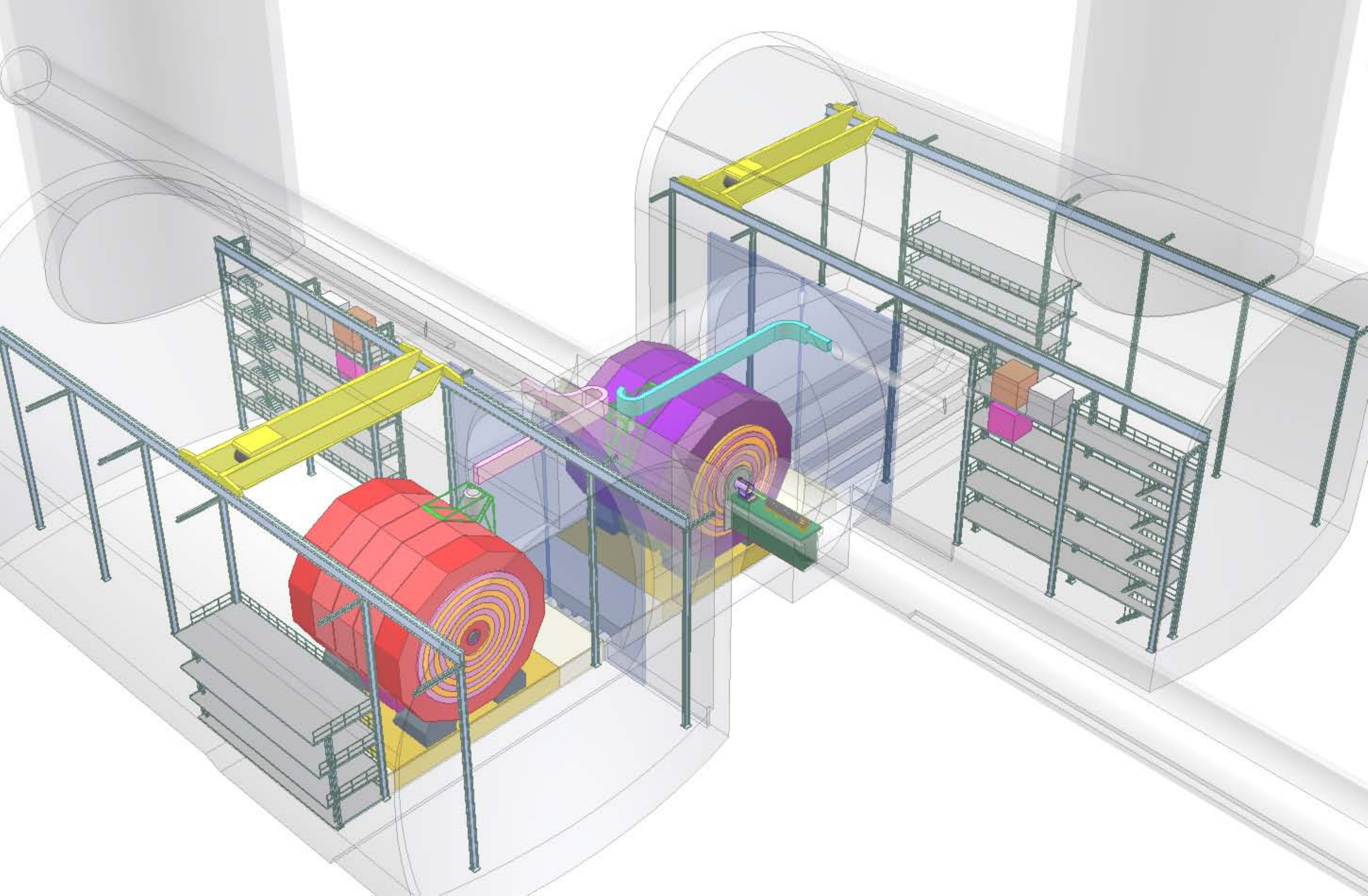}
\caption{Overall view of the interaction region at CLIC, with both detectors shown, one in the beam position and one in the maintenance position.}.
\label{fig:InteractionRegion}
\end{figure}

The beam delivery system at CLIC will have a single interaction region, with both detectors sharing the data taking periods in a``push-pull'' mode, with one detector in the beam position and one detector in a maintenance position in caverns on either side of the interaction point. The detectors, which each weigh around 12\,000 tonnes, are installed on independent, moveable platforms made from reinforced concrete. The close distance between the two detectors during beam operations imposes strict limits on the magnetic stray field of the solenoids, which is achieved by additional compensation coils in the case of CLIC\_ILD, and requires the detectors to be self-shielding to allow work on the detector in the garage position during beam operations. Figure \ref{fig:InteractionRegion} shows the overall layout of the CLIC interaction region.

Particularly challenging is the final focus system. The final focusing quadrupoles (QD0) are placed as close as possible to the IP, at a distance of less than 5 m, to achieve maximum luminosity. The small vertical size of the beam spot of 1 nm requires a stabilization of the QD0 position with an RMS of \mbox{0.15 nm} for frequencies above 4 Hz. This is achieved by supporting the QD0 with an active stabilization system from the accelerator tunnel. The QD0 will use a hybrid technology with permanent magnets and normal conducting coils, allowing to achieve a maximum gradient, while eliminating the need for cooling in the inner part, thereby reducing vibrations. The stability of QD0 is achieved by a combination of passive and active elements.

\section{Event Reconstruction and Background Mitigation}

The event reconstruction, and with it the whole design, of the CLIC detector concepts is based on particle flow, which reconstructs all visible particles in an event individually, making optimal use of the information available from the different detector subsystems. This applies in particular to the combination of tracking information and calorimetric measurements for the reconstruction of charged and neutral particles within dense hadronic jets. To evaluate the performance of the CLIC detectors in the challenging experimental environment presented by the high background rates, detailed simulation studies using full Geant4 \cite{Agostinelli:2002hh} models of the detectors and the inclusion of $\gamma\gamma \to  {\rm hadrons}$ background were performed in the framework of the CLIC CDR. The event reconstruction was performed using tracking and the PandoraPFA \cite{thomson:pandora} particle flow event reconstruction package. 

\begin{figure}[hbt]
\centering
 \includegraphics[width=0.49\linewidth]{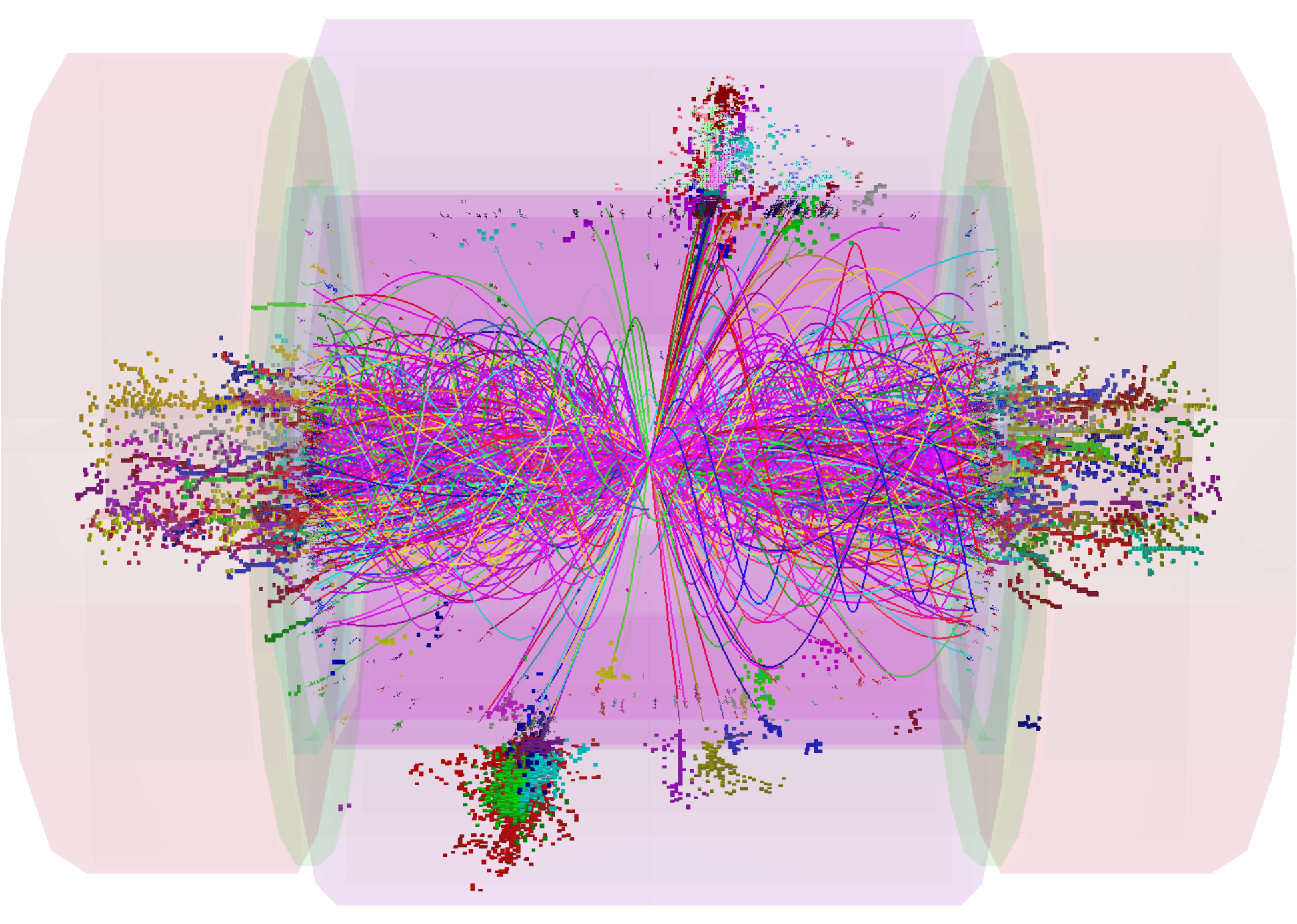} 
 \includegraphics[width=0.49\linewidth]{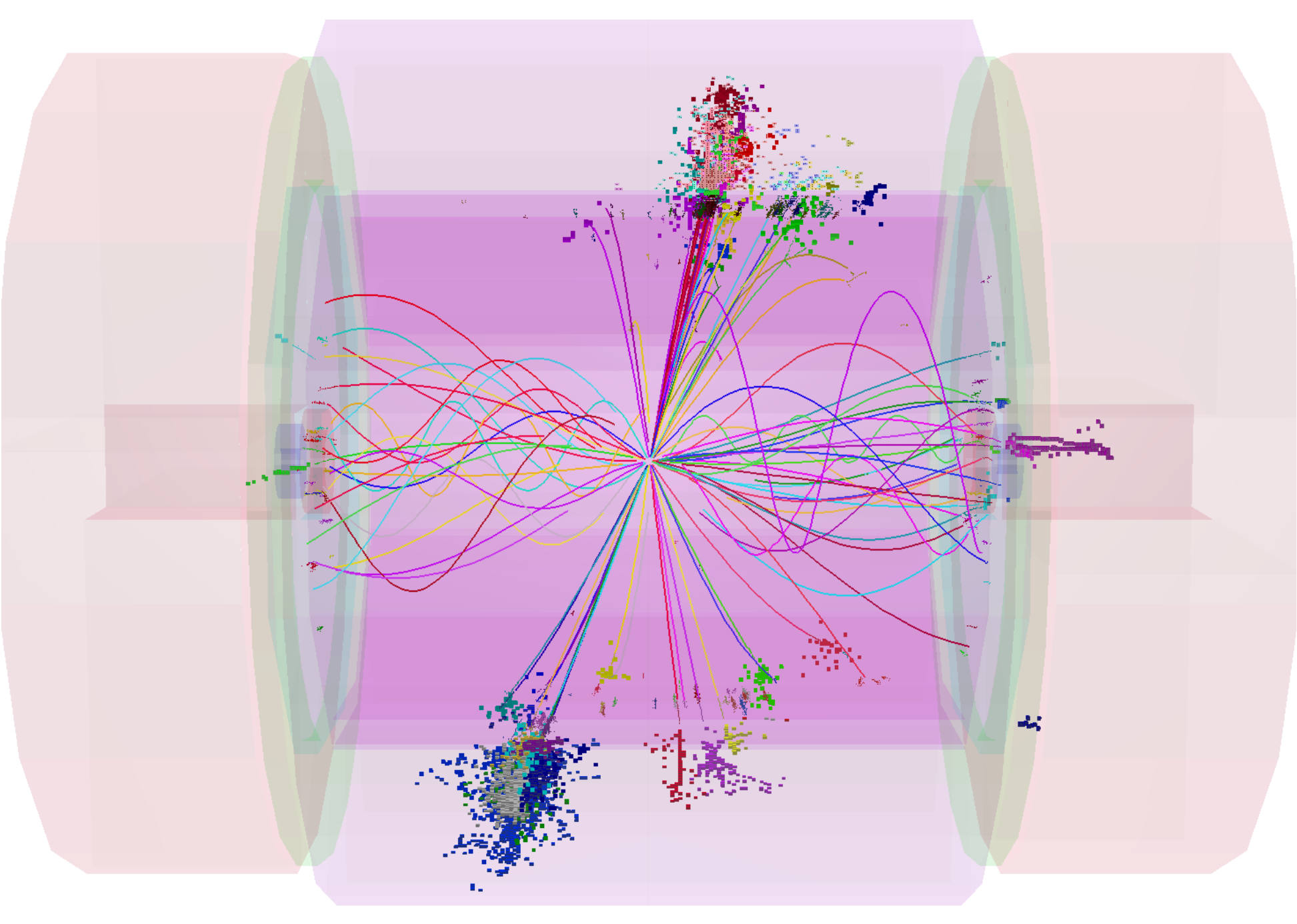}
 \caption{Event display of a 1 TeV di-jet event in the central detector region with 60 bunch crossings of $\gamma\gamma \to  {\rm hadrons}$ overlaid: (left) Event reconstruction without timing and other selection cuts. (right) Default selection cut.}
 \label{fig:EventDisplay}
\end{figure}

The impact of the background on the physics observables is reduced by introducing cuts on the level of reconstructed particles (referred to as particle flow objects (PFO)), selecting both on particle momentum and on timing. The time cuts depend on the transverse momentum and on the polar angle of the reconstructed particles, since background particles are found preferentially in the far forward regions and at low $p_T$. The selectivity of the timing cuts is driven by the precise time information available from the calorimeters. In the selection of particles, the flight time to the calorimeter surface is accounted for. For a complete bunch train, the additional energy deposited in the calorimeter from $\gamma\gamma \to  {\rm hadrons}$ background is around 19 TeV, which can be reduced to approximately 200 GeV by the selection cuts which lead to a signal loss of 0.5\% for typical physics events. For the full simulation study of physics processes, the equivalent of 60 bunch crossings of background have been added to physics events. Figure \ref{fig:EventDisplay} illustrates the impact of the background on a 1 TeV di-jet event, both without (left) and with (right) PFO selection cuts, showing that the background can very efficiently be rejected with appropriate cuts. 

\begin{figure}[!htb]
\centering
 \includegraphics[width=0.49\linewidth]{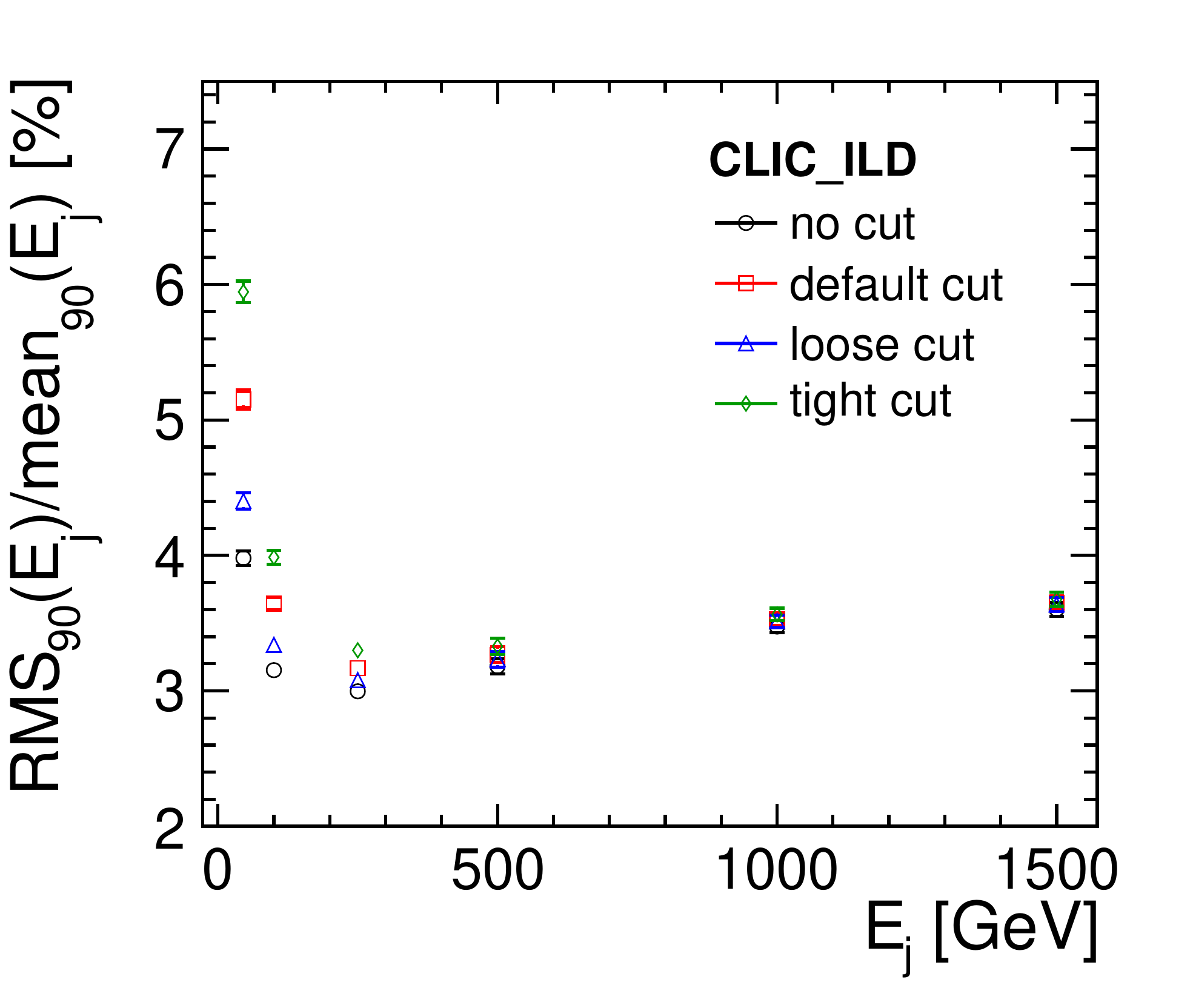} 
 \includegraphics[width=0.49\linewidth]{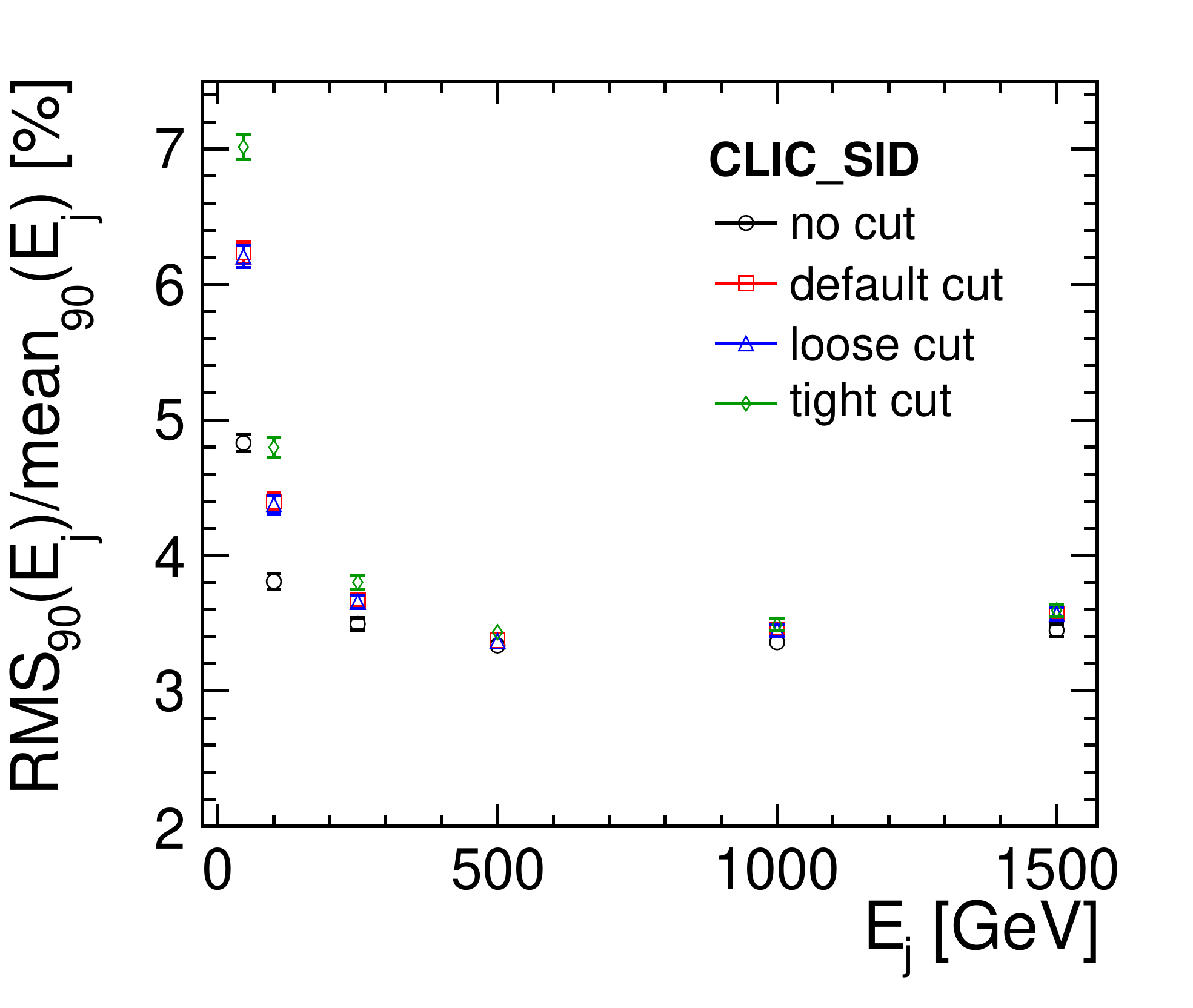}
 \caption{Impact of the timing cuts on the jet energy resolution as a function of the jet energy in di-jet events without detector backgrounds. Results are shown for the CLIC\_ILD ({\it left}) and CLIC\_SiD ({\it right}) detectors. The jet energy resolution is computed as the resolution of the total reconstructed energy multiplied by a factor of $\sqrt{2}$.}
 \label{fig:JetRes}
\end{figure}

At the same time, the jet energy resolution is not significantly deteriorated for high-energy jets which are typical in 3 TeV collisions. Only for energies below 200 GeV the impact of the cuts is noticeable, as shown in Figure \ref{fig:JetRes}, which shows the jet energy resolution for di-jet events fully simulated without $\gamma\gamma \to  {\rm hadrons}$ background, but with the application of various selection cuts. The reduced performance for low-energy jets affects mainly the physics at lower collision energies, where the background conditions are considerably less severe and the PFO selection cuts can be relaxed, recovering the original performance. 

In addition to these selection cuts, jet finding algorithms can contribute significantly to the rejection of background. The Durham $k_t$ algorithm used at LEP and for physics study at the ILC, which defines the distance between particles by the angle between them, proves to be very susceptible to additional background. By using an exclusive $k_t$ algorithm with a two-particle distance defined by pseudorapidity $\eta$ and azimuthal angle $\phi$, a metric commonly used in studies at hadron colliders, the impact of the mostly forward-going background particles is significantly reduced because of the stretching of the two-particle distance in that critical area.

\begin{figure}[!htb] 
	\centering
\includegraphics[width=0.5\textwidth]{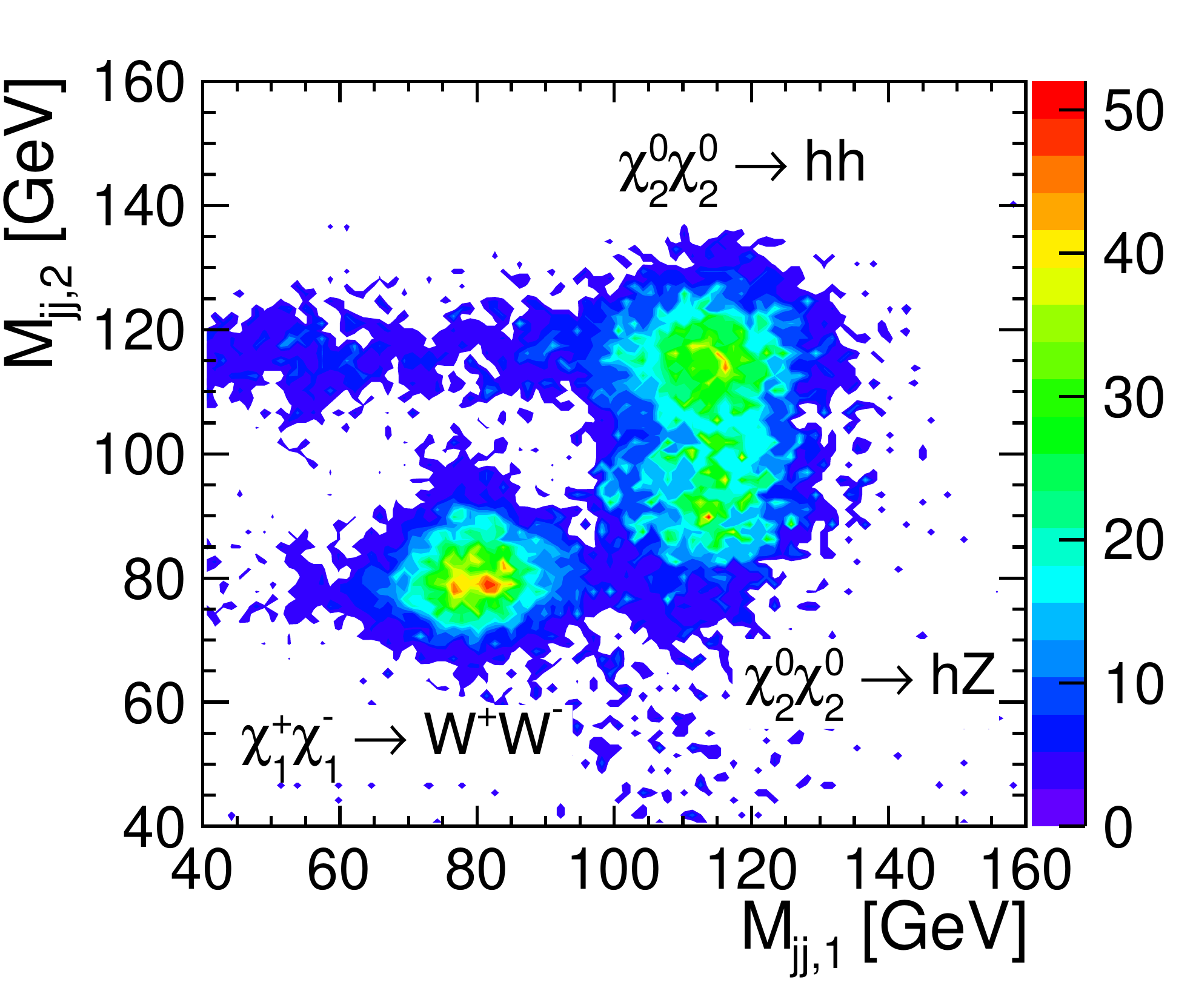}
\caption{Separation of heavy bosons in di-boson events with decay into all-hadronic final states from a full simulation of gaugino pair production in the CLIC\_SiD concept.}
\label{fig:Mass2D}
\end{figure}

A number of physics benchmark studies performed in the framework of the CLIC CDR have demonstrated that the CLIC detector concepts, combined with particle flow event reconstruction and timing cuts, are capable of precision measurements and the precise reconstruction of complicated final states in the challenging environment of a 3 TeV CLIC machine. Figure \ref{fig:Mass2D} illustrates the capability for efficient separation of different di-boson final states in all-hadronic decay topologies, originating from the pair production of neutralinos and charginos with masses in the 650 GeV range.

\section{Conclusions}

Two detector concepts have been developed for precision physics at the TeV scale with CLIC. These concepts are based on the validated detector designs for the ILC.  Both are multi-purpose detectors with high-precision vertex and main tracking and highly granular calorimetry inside a large solenoid, optimized for particle flow event reconstruction. The ILC concepts were adapted for the higher beam energy, in particular in the area of calorimetry, where a barrel hadron calorimeter with tungsten absorbers is used to provide increased depth while keeping the solenoid radius to technically feasible levels. To achieve the high luminosity required for the physics program, very high stabilization of the final focusing elements below the 1 nm level are necessary, requiring active stabilization techniques. A particular challenge at CLIC is the high rate of pair and $\gamma\gamma \to  {\rm hadrons}$  background, which necessitate increased radii for the beam pipe and vertex detectors, and make excellent time stamping capabilities necessary throughout the detector. With advanced reconstruction techniques using timing cuts on the single particle level, combined with appropriate jet finding in the physics analyses, the harsh background conditions can be controlled, paving the way for precision measurements of TeV-scale physics at CLIC.

\section*{Acknowledgments}
I would like to thank the organizers of TIPP2011 for the very interesting and stimulating conference. I thank my colleagues in the LCD project for help, advice and high quality material for this article.



%% References


\bibliographystyle{elsarticle-num}
\bibliography{CLICDetectors}






% \end{thebibliography}

\end{document}